\documentclass[aps,prb,showpacs,floatfix,twocolumn]{revtex4-1}

\usepackage{amsmath}
\usepackage{graphicx}
\usepackage{amstext,amsopn,amsmath,amssymb}
\usepackage{float}

\begin{document}

\title{Self-consistent rate theory for submonolayer surface growth\\ of
  multi-component systems}

\author{Mario Einax}
\email{mario.einax@uni-osnabrueck.de}
\affiliation{School of Chemistry, Tel Aviv University, Tel
Aviv 69978, Israel}
\affiliation{Fachbereich Physik, Universit\"at Osnabr\"uck,
Barbarastra{\ss}e 7, 49076 Osnabr\"uck, Germany}
\author{Philipp Maass}
\email{philipp.maass@uni-osnabrueck.de}
\affiliation{Fachbereich Physik, Universit\"at Osnabr\"uck,
Barbarastra{\ss}e 7, 49076 Osnabr\"uck, Germany}
\author{Wolfgang Dieterich}
\email{wolfgang.dieterich@uni-konstanz.de}
\affiliation{Fachbereich Physik, Universit\"at Konstanz, 78457
Konstanz, Germany}

\date{\today}

\begin{abstract}
  The self-consistent rate theory for surface growth in the
  submonolayer regime is generalized from mono- to multi-component
  systems, which are formed by codeposition of different types of
  atoms or molecules. As a new feature, the theory requires the
  introduction of pair density distributions to enable a symmetric
  treatment of reactions among different species. The approach is
  explicitly developed for binary systems and tested against kinetic
  Monte Carlo simulations. Using a reduced set of rate equations, only
  a few differential equations need to be solved to obtain good
  quantitative predictions for island and adatom densities, as well as
  densities of unstable clusters.
\end{abstract}


\maketitle

\section{Introduction}
\label{sec:intro}
Growth of solid structures on surfaces, induced by atomic or molecular
deposition, has become a widely applied method for generating
materials of nanoscale dimensions.\cite{Brune:1998,
  Ratsch/Venables:2003, Michely/Krug:2004, Evans/etal:2006,
  Kuehnle:2009, Hlawacek/Teichert:2013, Rahe/etal:2013} The resulting
clusters or thin films are often metastable and their structure
depends on kinetics rather than thermodynamics. Understanding and
control of such growth processes are prerequisites for designing
nanomaterials of practical use.  Multi-component systems are
particularly promising in this respect because of their larger
structural variability compared to single-component
systems.\cite{Einax/etal:2013} In the submonolayer growth regime
one-monolayer islands can act as seeds for 3D structures that emerge
in later stages of growth.\cite{Koerner/etal:2011, Albrecht/etal:2002,
  Liscio/etal:2010, Einax/etal:2007b}

Island nucleation and submonolayer growth of binary systems, driven by
co-deposition of two species $A$ and $B$, has recently been
investigated by using rate equations and kinetic Monte-Carlo (KMC)
simulations.\cite{Einax/etal:2007a} Generalized relations were
established that describe the scaling of stable island densities with
the partial fluxes $F_\alpha$ ($\alpha=A$ or $B$), adatom diffusion
coefficients $D_\alpha$ and mutual binding energies
$E_{\alpha\beta}$. Simulations also showed that island density data,
when combined for different compositions, enable to extract
microscopic parameters for mixed systems.\cite{Einax/etal:2009} Of
particular value is the possibility to determine the binding energy
$E_{AB}$ between unlike atoms in the presence of a surface.

In the rate equations for submonolayer growth,\cite{Venables:1973,
  Venables/etal:1984, Venables:2000} capture numbers $\sigma_s$ appear
as parameters, which determine the attachment rate of diffusing
adatoms to islands of size $s$. Already in the single-component case
it is known that for a quantitative description of island densities as
a function of coverage $\Theta$, it is essential to deal with
effective capture numbers
$\sigma_s(\Theta,\Gamma)$.\cite{Gibou/etal:2003, Popescu/etal:2001,
  Koerner/etal:2010, Koerner/etal:2012} Their dependence on the
coverage $\Theta$ and the ``$D/F$-ratio'', $\Gamma=D/F$, reflects the
fact that the efficiency of an island of size $s$ to capture adatoms
is affected by the shielding by other islands in its
neighborhood. Within a mean-field description of these shielding
effects, a central $s$-sized island is thought to be embedded in an
effective medium, characterized by an absorption length $\xi$ for the
adatoms. This length describes the capture efficiencies of all islands
in an averaged manner. As the rate of capture by the central island is
determined by the $\xi$-dependent adatom density profile in its
vicinity, one arrives at a self-consistency condition for
$\sigma_s$. Originally, this self-consistent theory was formulated for
diffusion-limited irreversible growth.\cite{Bales/Chrzan:1994} Later
it has been extended to include detachment
kinetics,\cite{Bales/Zangwill:1997, Popescu/etal:1998} and to examine
capture numbers in the presence of cluster
diffusion\cite{Hubartt/etal:2011} and adsorbate
interactions.\cite{Venables/Brune:2002, Ovesson:2002}

Our goal here is to generalize the self-consistent theory of
diffusion-limited growth to multi-component systems. In order to
obtain capture numbers which are symmetric under the exchange of
species, it is needed to introduce pair distribution functions. The
treatment will be focuses on binary systems, where trimers and larger
islands are stable irrespective of composition, whereas the stability
of dimers $AA$, $AB$, and $BB$ is allowed to be composition dependent.
Generalizations are discussed in Sec.~\ref{sec:conclusion}.

\section{Rate equations for binary systems}
\label{sec:rate_eqs}
Following earlier work\cite{Einax/etal:2007a, Dieterich/etal:2008} we
start out from rate equations for island densities in a system of two
species $A$ and $B$. For simplicity, we will speak about $A$ and $B$
``atoms'', but these could be also molecules, if their geometrical
arrangement with respect to the substrate topology does not play an
essential role for the time evolution of island densities. The $A$ and
$B$ species are assumed to be deposited as adatoms (no cluster
deposition) and to be mobile on the surface. They undergo nucleation
and dissociation reactions among themselves, and they attach to and
detach from already formed islands of larger size. These larger
islands are considered to be immobile. The coverage $\Theta$ is
supposed to be small enough so that coalescence of islands can be
neglected. Direct impingement of arriving atoms onto already existing
islands and desorption processes are neglected, or they may be taken
into account by introducing properly re-scaled fluxes. Furthermore, we
limit our discussion to cases where the largest unstable islands are
composed of not more than two atoms. Then the time evolution of adatom
densities $n_\alpha$, $\alpha=A,B$, is given by
\begin{align}
\label{eq:monomer_densities}
\frac{d n_\alpha}{dt} =& F_\alpha - 2 D_\alpha \sigma_1^{\alpha \alpha} n_\alpha^{2} -
\left( D_A + D_B \right) \sigma_1^{AB} n_A n_B \\
&\hspace*{-2em} - D_\alpha n_\alpha \sum_{s\geq 2} \sigma_{s}^{\alpha} n_{s} +
{K}_2^{AB} n_{AB} + 2 {K}_{2}^{\alpha \alpha} n_{\alpha \alpha} \, . \nonumber
\end{align}
Positive contributions to~(\ref{eq:monomer_densities}) arise from the
partial fluxes $F_{\alpha}=x_{\alpha}F$, with $x_{\alpha}$ the
fraction of $\alpha$-atoms and $F=F_A+F_B$ the total flux, and from
the decay of the different kinds of dimers with densities $n_{\alpha
  \beta}$. Negative contributions refer to the formation of dimers and
attachment of adatoms to $s$-sized islands.  Note that the diffusion
coefficient for the relative motion of $A$ and $B$ is $D_A+D_B$. In
the sum over $s$, the term $s=2$ involves
$n_2=n_{AA}+n_{AB}+n_{BB}$. The rate equations for dimer densities are
\begin{align}
  \frac{dn_{\alpha \alpha}}{dt} &=
D_{\alpha} \sigma_1^{\alpha \alpha} n_{\alpha}^2\nonumber\\
 &{}-\bigl(\sum_\beta D_\beta \sigma_2^{\beta} n_\beta \bigr)
 n_{\alpha \alpha}
- K_2^{\alpha \alpha} n_{\alpha \alpha}\, ,
 \label{eq:Dimer_densities}
\end{align}
\begin{align}
  \frac{dn_{AB}}{dt} &= (D_A +
D_B) \sigma_1^{AB} n_A n_B \nonumber\\
 &{}-\bigl(\sum_\beta D_\beta \sigma_2^{\beta} n_\beta \bigr) n_{AB}
- K_2^{AB} n_{AB}\, .
 \label{eq:Dimer_AB_density}
\end{align}
The upper indices in the capture numbers $\sigma_1^{\alpha\beta}$,
$\sigma_s^\alpha$ and decay rates $K_2^{\alpha\beta}$ serve to
distinguish the types of adatoms that are involved in a reaction. The
$\sigma_1^{\alpha\beta}$ and $K_2^{\alpha\beta}$, respectively,
refer to formation and dissociation of an $\alpha\beta$-dimer.  The
$\sigma_s^{\alpha}$, $s\ge2$, refer to the capture of an $\alpha$
adatom by an island composed of $s$ atoms.  The geometry of such
island is represented by a circular shape (formation of compact
islands) with radius $\mathcal{R}_s=s^{1/2}\mathcal{R}_1$, where
$\mathcal{R}_1$ is the adatom radius.  Since we allow
composition-dependent (``mixed'') dimer stabilities, some of the decay
rates can be zero. For the purpose of calculating $n_{\alpha}$,
$n_{\alpha \beta}$ and the total density of stable islands, $N$, it
appears sufficient to ignore any further composition dependencies of
parameters beyond those given in Eqs.~(\ref{eq:Dimer_densities}) and
(\ref{eq:Dimer_AB_density}).

The densities of islands with $s>2$ evolve according to
\begin{align}
\label{eq:s-cluster}
\frac{d n_s}{dt} &= \sum_\alpha D_\alpha n_\alpha
\left( \sigma_{s-1}^{\alpha} n_{s-1}
- \sigma_{s}^{\alpha} n_{s} \right) \, .
\end{align}

\section{Irreversible growth}
\label{sec:i1}
In the self-consistent rate theory, analytical expressions for the
capture numbers and decay rates are derived by introducing an
effective medium that describes adatom capture in an averaged manner
by an absorption length $\xi$. For binary systems, the effective
medium is characterized by two different absorption lengths
$\xi_\alpha$ for the two adatom species. To define $\xi_\alpha$, the
evolution equations (\ref{eq:monomer_densities}) for monomer densities
with zero decay terms ($i=1$) are rewritten as
\begin{align}
\label{eq:MFRE_with_tau}
\frac{d n_\alpha}{dt} &= F_\alpha -\frac{1}{\tau_{\alpha}} n_\alpha \,,
\end{align}
where $\tau_\alpha^{-1}=D_\alpha/\xi_\alpha^2$ is the reaction rate of
$\alpha$ adatoms in the effective medium, and
\begin{align}
\label{eq:xi_alpha}
\xi_\alpha^{-2}
&= \sum_\beta\left(1-\delta_{\alpha \beta}\right)\sigma_1^{\alpha
  \beta}
\left(1+\frac{D_\beta}{D_\alpha}\right) n_\beta
\nonumber \\
& +2 \sigma_1^{\alpha \alpha} n_\alpha + \sum_{s\geq 2} \sigma_s^{\alpha} n_s\,.
\end{align}
Deposition, diffusion and absorption of adatoms within the effective
medium are described by local densities $\tilde{n}_{\alpha}
\left(\textbf{r}\right)$ with $n_\alpha=\int_V d^2 r\,
\tilde{n}_\alpha (\textbf{r})/V$, where $V$ is the two-dimensional
volume (surface area). These satisfy
\begin{align}
\label{eq:n_tilde_alpha}
\frac{\partial \tilde{n}_\alpha}{\partial t}
&= F_\alpha + D_\alpha \Delta \tilde{n}_\alpha -
\frac{1}{\tau_\alpha} \tilde{n}_\alpha \, .
\end{align}
In the mono-component case, one would have just one equation of this
type, and by supplementing this with appropriate boundary conditions,
the stationary density profiles of adatoms around islands with radius
$R_s$ can be calculated and the total adatom flux to the islands
identified with the corresponding capture terms in
Eqs.~(\ref{eq:monomer_densities}). This procedure yields
self-consistent analytical expressions for the capture numbers and
decay rates in the mono-component case.

For binary (multi-component) systems, the reaction between unlike
adatoms needs a refined treatment. This has the following reason: In a
naive extension of the monocomponent case, the $B$ adatom density
around an $A$ adatom would by characterized by an absorption length
$\xi_B$, and the $A$ adatom density around a $B$ adatom by an
absorption length $\xi_A$.  However, the shape of both profiles is
given by the pair density ${n}_{AB}(\textbf{r},\textbf{r}')$ of $A$
and $B$ adatoms and hence the profiles must be characterized by the
same capture length (if inversion symmetry holds). In fact,
introducing the pair distribution function\cite{Hansen/McDonald:1986,
  Kotomin/Kuzovkov:1996}
\begin{align}
\label{eq:pair_distr_func}
G_{AB}(\textbf{r}) &= \frac{1}{V} \hspace*{-0.1cm}
\int_{V} \hspace*{-0.15cm} d^2r' \hspace*{-0.15cm}
\int_{V} \hspace*{-0.15cm} d^2r'' \,
{n}_{AB}\left(\textbf{r}',\textbf{r}''\right)
\delta\left(\textbf{r}-(\textbf{r}'-\textbf{r}'')\right)
\end{align}
allows one to treat unlike adatoms in a symmetric way, resulting in a
symmetric expression for $\sigma_1^{AB}$. $G_{AB}(\textbf{r})$ is the
number of pairs of $A$ and $B$ adatoms at distance $\textbf{r}$ per
area. Let us note that the approach based on pair distribution
functions is well known in the kinetic theory of bimolecular
chemical reactions.\cite{Waite:1957, Kotomin/Kuzovkov:1996} In our
context, spatial correlations between adatoms for relative distances
larger than the contact distance $R_1=2\mathcal{R}_1$ play no role so
that
\begin{align}
\label{eq:pair_density}
n_{AB}(\textbf{r},\textbf{r}')&=
\tilde{n}_A(\textbf{r})\tilde{n}_B(\textbf{r}')\,,
\qquad |\textbf{r}-\textbf{r}'|>R_1\,.
\end{align}
Combination with (\ref{eq:n_tilde_alpha}) yields an expression for the
time derivative of $n_{AB}(\textbf{r},\textbf{r}')$. Subsequent
multiplication by $\delta(\textbf{r}-(\textbf{r}'-\textbf{r}''))$ and
integration over all $\textbf{r}'$ and $\textbf{r}''$ gives
\begin{align}
\label{eq:BWGL_G}
\frac{\partial G_{AB}(\textbf{r})}{\partial t}
&= F_A n_A + F_B n_B  + \left( D_A + D_B \right)
\Delta G_{AB} (\textbf{r}) \nonumber\\
&-\left( \frac{1}{\tau_A} + \frac{1}{\tau_B} \right)
G_{AB}(\textbf{r})\, .
\end{align}
Subtracting $d(n_An_B)/dt$ with the help of
Eqs.~(\ref{eq:MFRE_with_tau}) and going over to the quasi-stationary
limit, we obtain
\begin{align}
\label{eq:BWGL_G_steady_state_cond}
\left( D_A + D_B \right) \Delta G_{AB} (\textbf{r})
&= \left( \frac{1}{\tau_A} + \frac{1}{\tau_B} \right)
\left(G_{AB} (\textbf{r}) - n_A n_B\right).
\end{align}
Alternatively,
\begin{align}
\label{eq:BWGL_G_steady_state}
\Delta G_{AB}(\textbf{r}) - \frac{1}{\xi_{\rm eff}^2}
\left(G_{AB} (\textbf{r}) - n_A n_B\right) &= 0 \, ,
\end{align}
where we introduced the effective absorption length
\begin{align}
\label{eq:effective_length}
\xi_{\rm eff}^{-2} &= \frac{1}{D_A+D_B} \left( \frac{D_A}{\xi_A^2} +
  \frac{D_B}{\xi_B^2} \right) \,,
\end{align}
which is a weighted average of $\xi_{\alpha}^{-2}$, with weighting
factors $D_{\alpha}/(D_A+D_B)$.

For $i=1$, implying complete absorption at contact, and assuming
isotropy, the boundary conditions to
Eq.~(\ref{eq:BWGL_G_steady_state}), are
\begin{align}
\label{eq:BC_G_AB}
G_{AB}(r) &\rightarrow \left\{
\begin{array}{ll}
n_A n_B & \qquad r \rightarrow \infty \, ,\\
0       & \qquad r \rightarrow R_1 \,,
\end{array}
\right.
\end{align}
where $r=|\textbf{r}|$ and we have replaced $G_{AB}(\textbf{r})$ by
$G_{AB}(r)$. The solution of Eq.~(\ref{eq:BWGL_G_steady_state}) with
the boundary conditions in Eq.~(\ref{eq:BC_G_AB}) is
$G_{AB}(r)=n_An_B([1-\mathcal{K}_0(r/\xi_{\rm
  eff})/\mathcal{K}_0(R_1/\xi_{\rm eff})]$, where $\mathcal{K}_{\nu}$
is the modified Bessel function of order $\nu$.

To obtain the reaction rate, we first select reactions along a particular
direction $\hat{\textbf{r}}=\textbf{r}/|\textbf{r}|$,  $\textbf{r}$ being
the relative coordinate between an $A$ and $B$ atoms right before contact.
The corresponding rate is given by
\begin{align}
\label{eq:reaction_rate}
I(\hat{\textbf{r}}) &= \lim_{|\textbf{r}| \rightarrow R_1}
\frac{1}{V} \int d^2r' \int d^2r'' \,
\hat{\textbf{r}} \cdot
\left[ \textbf{j}_B (\textbf{r}'') \tilde{n}_A(\textbf{r}')
\right.\nonumber\\
&\hspace*{5em}\left. - \textbf{j}_A (\textbf{r}') \tilde{n}_B(\textbf{r}'')
\right] \, \delta (\textbf{r}-(\textbf{r}'-\textbf{r}''))\,,
\end{align}
where
$\textbf{j}_\alpha(\textbf{r})=-D_\alpha\nabla\tilde{n}_\alpha(\textbf{r})$.
Substituting this expression into (\ref{eq:reaction_rate}) and using
(\ref{eq:pair_distr_func}), we can reexpress (\ref{eq:reaction_rate})
as $I(\hat{\textbf{r}})=(D_A+D_B)(\partial G_{AB}/\partial r)|_{R_1}$.
After integration along the boundary at $r=R_1$, we obtain the total
number of reactions per second and per unit area, which is identified
with the corresponding term in the original rate equations, $(D_A+D_B)
\sigma_1^{AB} n_A n_B$. Thus, we obtain
\begin{align}
\label{eq:sigma_AB_flux}
\sigma_1^{AB} &= 2 \pi R_1 \frac{1}{n_A n_B}
\left( \frac{\partial G_{AB}}{\partial r}\right)_{R_1} \, .
\end{align}

Evidently, this result for $AB$-capture in a binary system has a
structure analogous to the self-consistent capture number $\sigma_1$
for a one-component system of overall adatom density $n$ and diffusion
coefficient $D$. That situation and the present one can be mapped onto
each other by $n \leftrightarrow n_A n_B$; $2D \leftrightarrow
D_A+D_B$; $\tilde{n} (\textbf{r}) \leftrightarrow G_{AB}(\textbf{r})$
for the local densities in the SCF-treatment, and $\xi \leftrightarrow
\xi_{\rm eff}$, where $\xi_{\rm eff}$ was defined by
(\ref{eq:effective_length}).  Hence we can immediately translate known
results for one-component systems to the present case, to obtain
\begin{align}
\label{eq:sigma_1_AB}
\sigma_1^{AB} &= 2 \pi \frac{R_1}{\xi_{\rm eff}}
\frac{\mathcal{K}_1 \left( R_1/\xi_{\rm eff} \right)}{\mathcal{K}_0
\left( R_1/\xi_{\rm eff} \right) } \, .
\end{align}

The $\sigma_1^{\alpha \alpha}$ are obtained by introducing the pair
correlation function $G_{\alpha \alpha}(r)$ for like particles and
repeating the above steps. For $\sigma_1^{\alpha \alpha}$ we recover
the form (\ref{eq:sigma_1_AB}) with $\xi_{\rm eff}$ replaced by
$\xi_{\alpha}$.  Moreover, we need $\sigma_s^{\alpha}$ for $s\ge2$.
Since islands with $s\ge2$ do not move, the result is again equivalent
to (\ref{eq:sigma_1_AB}), where one type of adatoms has zero diffusion
coefficient. For example, $\sigma_s^A$ is given by
(\ref{eq:sigma_1_AB}) with $D_B=0$, hence $\xi_{\rm eff}$=$\xi_A$, and
$R_1$ replaced by $R_s=\mathcal{R}_s+\mathcal{R}_1$. Clearly, our
treatment also covers one-component systems through the limit where
$A$ and $B$ atoms become indistinguishable.

\section{Decay processes}
\label{sec:decay}
In this section we extend the above scheme to include detachment
processes.  First, we focus on unstable $AB$-dimers, characterized by
some finite binding energy $E_{AB}$.\cite{Einax/etal:2009} This
situation can be incorporated into the treatment of Sec.~\ref{sec:i1}
by a modification of the boundary condition
(\ref{eq:BC_G_AB}). Consider detachment and re-attachment reactions
between an $A$ and $B$ adatom.  Within a lattice model and $E_{AB}$ a
nearest neighbor binding energy, the bound state corresponds to an
$AB$-pair located on nearest neighbor sites, whereas in the detached
state the $A$ and $B$ adatoms are separated by one vacant site.  By
$n_{AB}$ and $n^*_{AB}$ we denote the densities of bound and
detached states of this type. Assuming local equilibrium, both
densities are related by
\begin{align}
\label{eq:local_equilibrium}
 n^*_{AB} &= n_{AB}\,\mu_{AB}\,\exp\left(-E_{AB}/k_{\rm B}T\right) \, .
\end{align}
The factor $\mu_{AB}$ is determined by the degeneracies of the bound
and dissociated states in a circularly averaged description, and
depends on the geometry of $A$ and $B$ adsorption sites on the
surface. We do not go into the underlying counting problem for
specific lattice geometries,\cite{Popescu/etal:1998} but merely treat
$\mu_{AB}$ as a parameter.\cite{Einax/etal:2007a} Writing $\mu_{AB}
\exp\left(-E_{AB}/k_{\rm B} T\right)=\kappa_{AB}$ we arrive at the
local equilibrium boundary condition
\begin{align}
\label{eq:BC_i2}
G_{AB}(r) \rightarrow  \kappa_{AB}\,n_{AB} \,, \qquad r \rightarrow R_2\, .
\end{align}
As before, see Eq.~(\ref{eq:BC_G_AB}), $G_{AB}(r)\rightarrow n_A n_B$
as $r \rightarrow \infty$.  Solving Eq.(\ref{eq:BWGL_G_steady_state})
for these boundary conditions yields $G_{AB}(r)=n_An_B([1-\zeta
\mathcal{K}_0(r/\xi_{\rm eff})/\mathcal{K}_0(R_1/\xi_{\rm eff})]$ with
$\zeta=(1-\kappa_{AB}n_{AB}/n_An_B)$.

The total reaction rate can be then written
as\cite{Bales/Zangwill:1997}
\begin{align}
\label{eq:total_reaction_rate}
I_{\rm tot}&=I_{\rm capture}-I_{\rm decay}\, ,
\end{align}
where $I_{\rm capture}=(D_A+D_B) \sigma_1^{AB} n_A n_B$ is defined
with $\sigma_1^{AB}$ from Eq.~(\ref{eq:sigma_AB_flux}), and
\begin{align}
\label{eq:decay_current}
I_{\rm decay}&=\frac{n_{AB}}{n_A n_B}\kappa_{AB} I_{\rm capture} \, .
\end{align}
Identification with the corresponding decay term $K_2^{AB} n_{AB}$ in
Eq.~(\ref{eq:Dimer_AB_density}), we find
\begin{align}
\label{eq:decay_rate_K2_AB}
K_2^{AB}=(D_A+D_B) \kappa_{AB} \sigma_1^{AB} \, .
\end{align}
%
\begin{figure*}[tbp!]
\includegraphics[width=0.4\textwidth,clip=,]{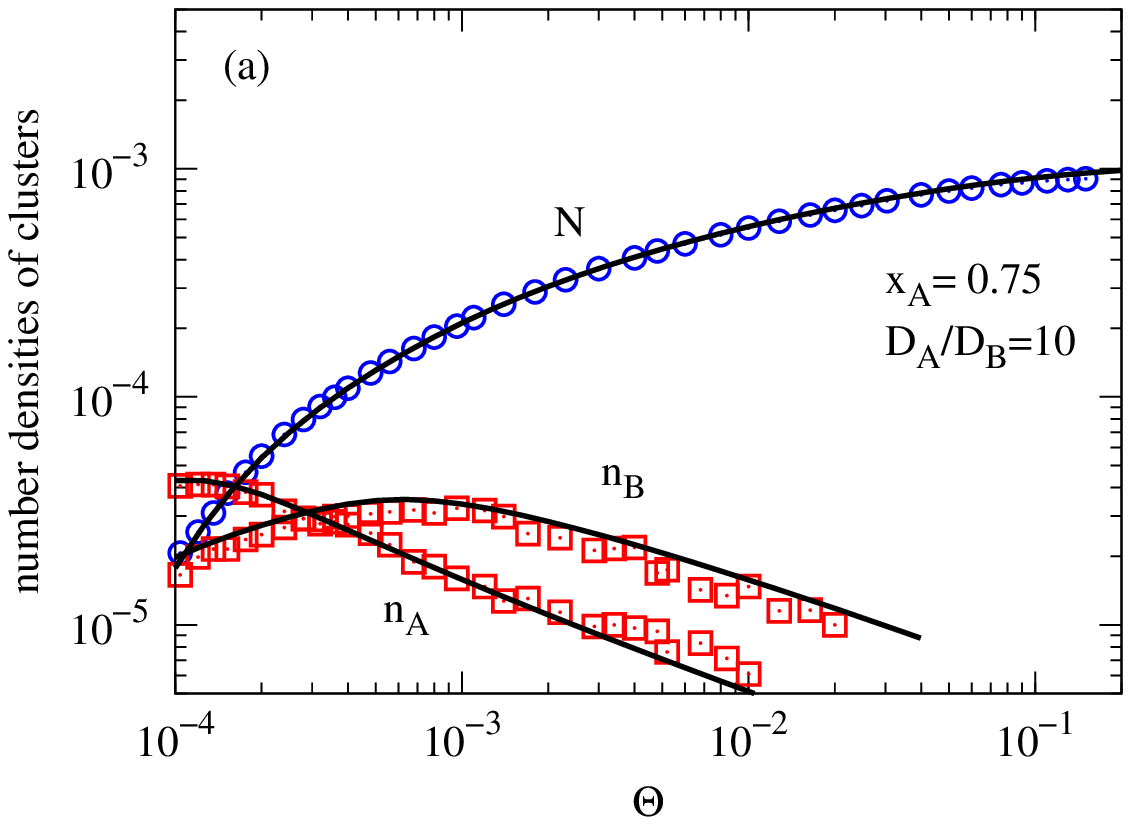}
\includegraphics[width=0.4\textwidth,clip=,]{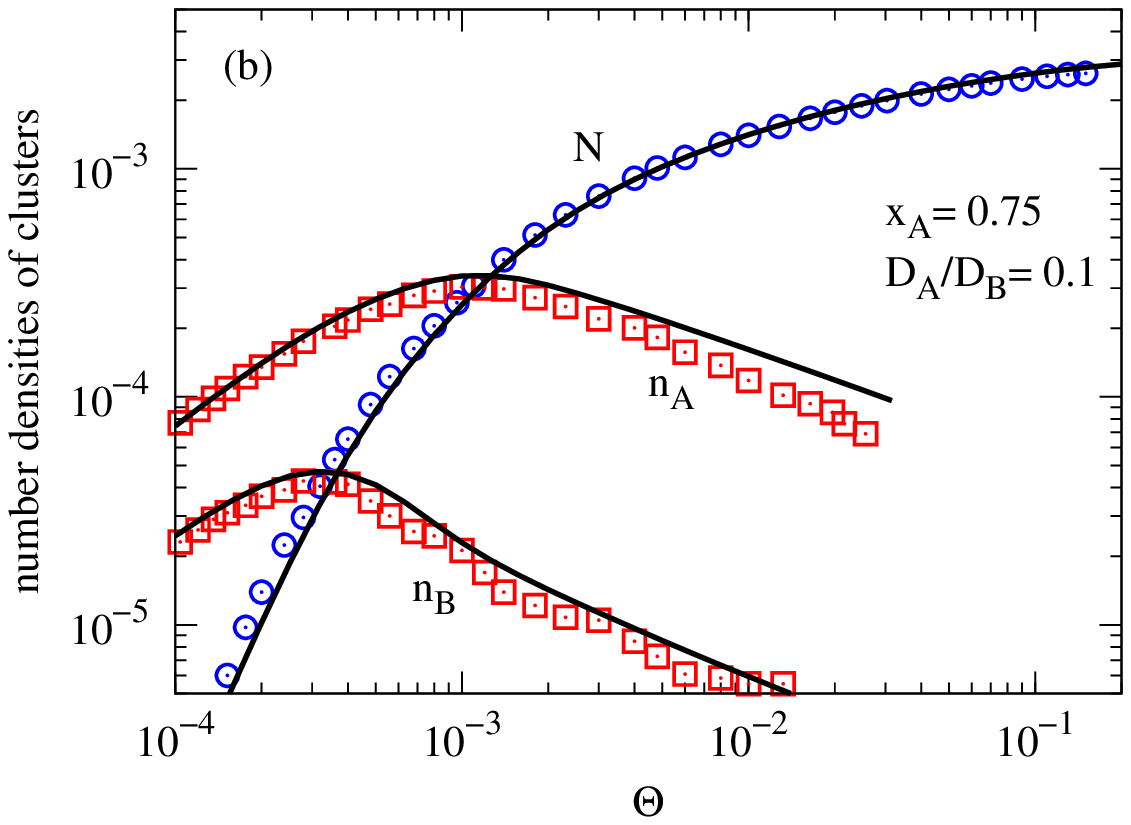}
\includegraphics[width=0.4\textwidth,clip=,]{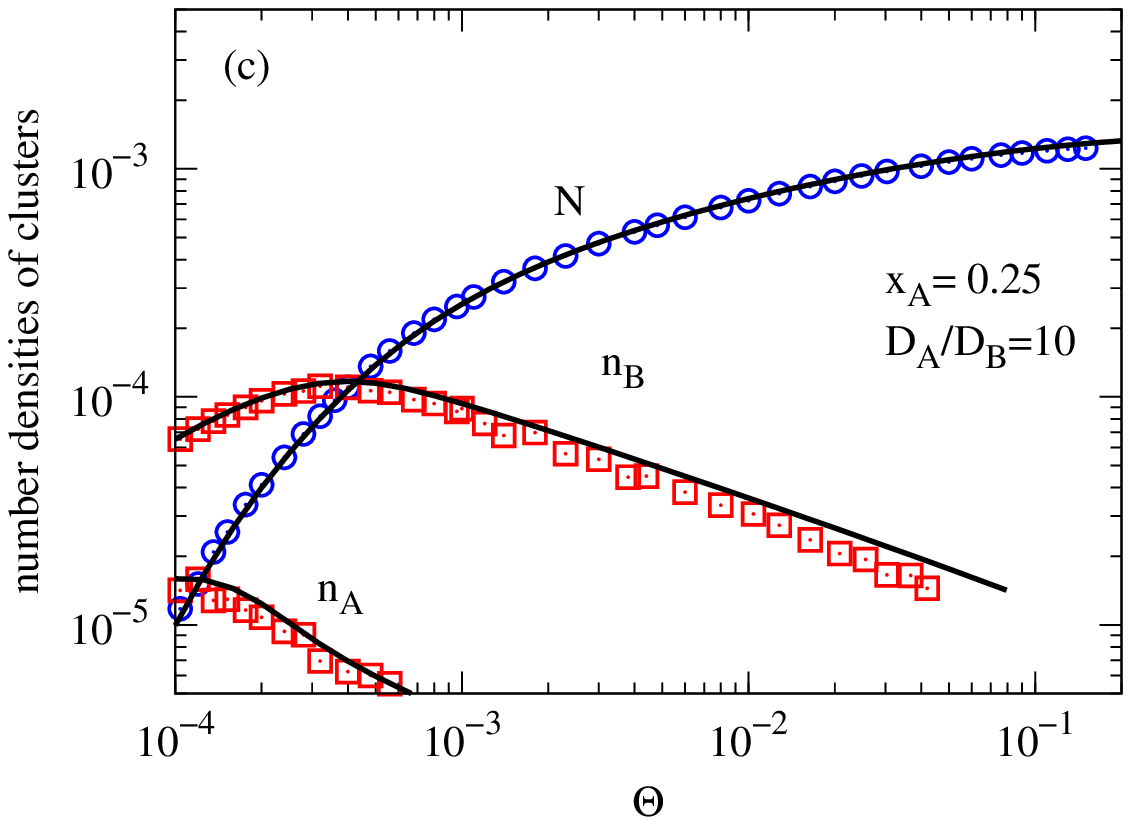}
\includegraphics[width=0.4\textwidth,clip=,]{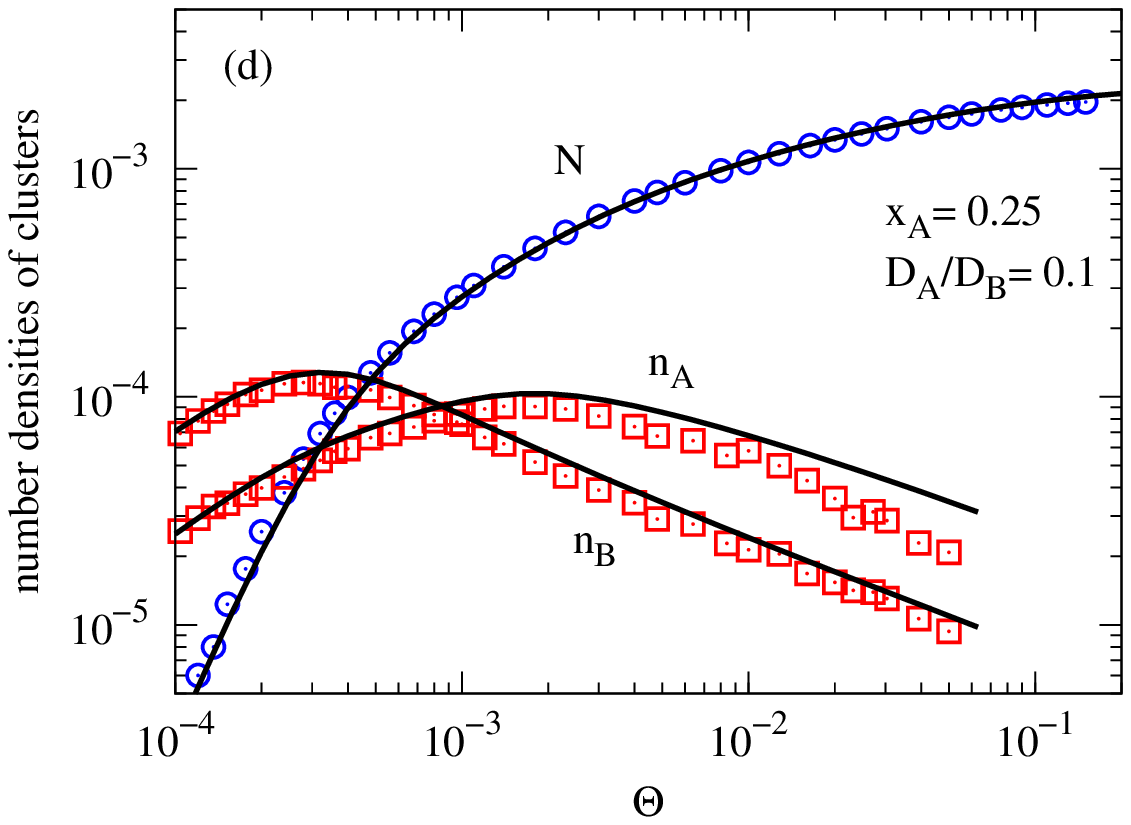}
\caption{(Color online) Number densities of $A$ and $B$ adatoms, and
  stable islands as a function of the coverage $\Theta= F t$ for $i=1$
  and various combinations of $x_A$ and $D_A/D_B$. Results from the
  self-consistent rate theory are given by solid lines.}
\label{fig:fig1}
\end{figure*}
%
In the same way we obtain
\begin{align}
\label{eq:decay_rate_K2_alpha}
K_2^{\alpha\alpha}=2D_{\alpha} \kappa_{\alpha \alpha} \sigma_1^{\alpha \alpha}
\end{align}
with $\kappa_{\alpha\alpha}=\mu_{\alpha\alpha}
\exp(-E_{\alpha\alpha}/k_{\rm B} T)$.  Again, the degeneracy factors
$\mu_{\alpha \alpha}$ are treated as parameters.

Note that when we use these results for the self-consistent capture
and decay numbers in the two-component Walton
relations,\cite{Einax/etal:2007a} $(D_A+D_B) \sigma_1^{AB} n_A n_B
\simeq K_2^{AB} n_{AB}$, it follows that $I_{\rm capture} \simeq
I_{\rm decay}$. This is consistent with the quasi-stationarity
assumption underlying the Walton relations,\cite{Walton:1962} which
implies that the capture and decay rates nearly balance. Let us
further note that reaction barriers for formation and dissociation of
dimers can also be incorporated in the treatment. They lead to a
modification of the boundary condition (\ref{eq:BC_i2}), corresponding
to a partially reflecting boundary, sometimes called ``radiative
boundary condition''.\cite{Waite:1957,Bales/Zangwill:1997,Kandel:1997}

\begin{figure}[!htbp] 
\includegraphics[width=0.4\textwidth,clip=,]{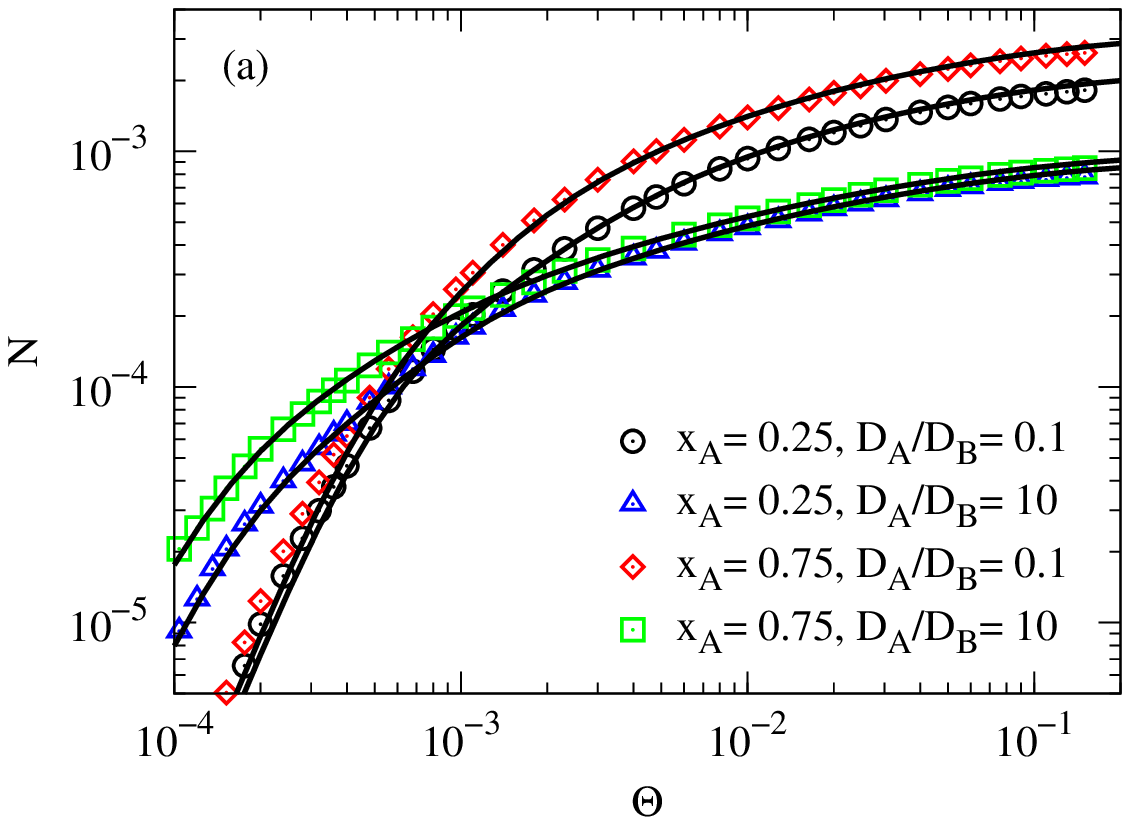}
\includegraphics[width=0.4\textwidth,clip=,]{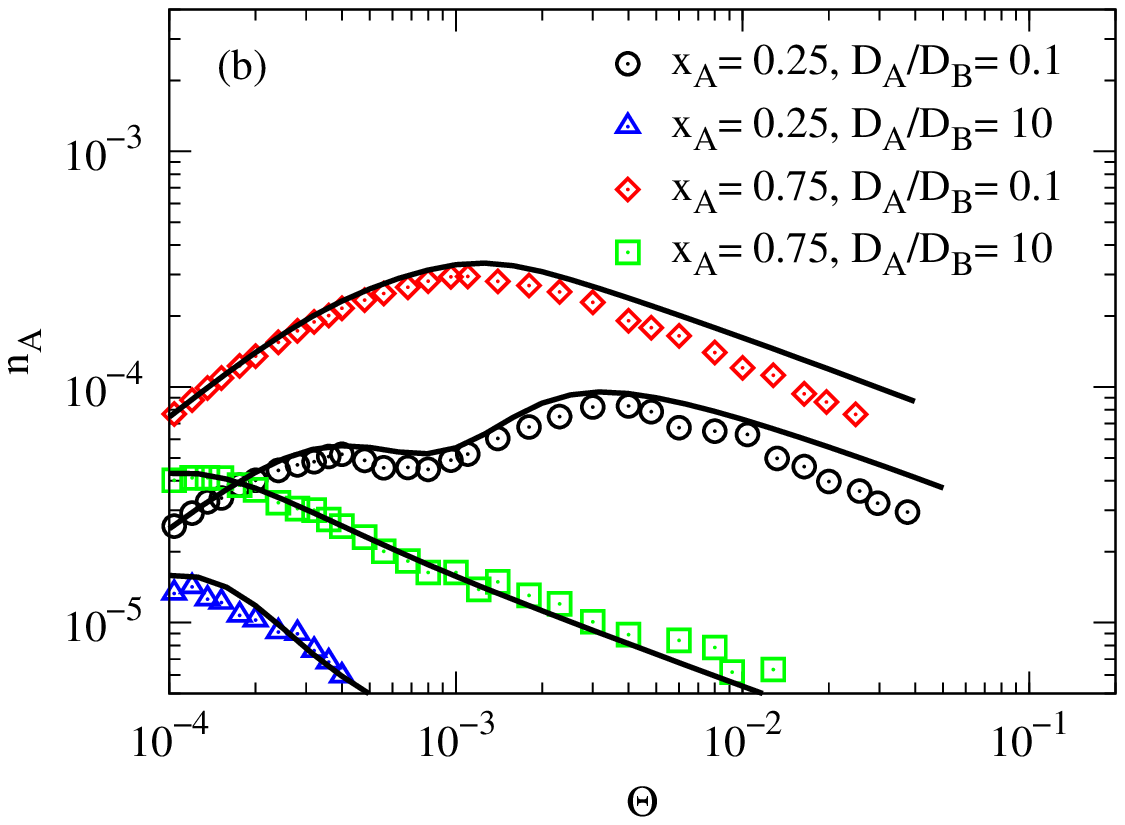}
\includegraphics[width=0.4\textwidth,clip=,]{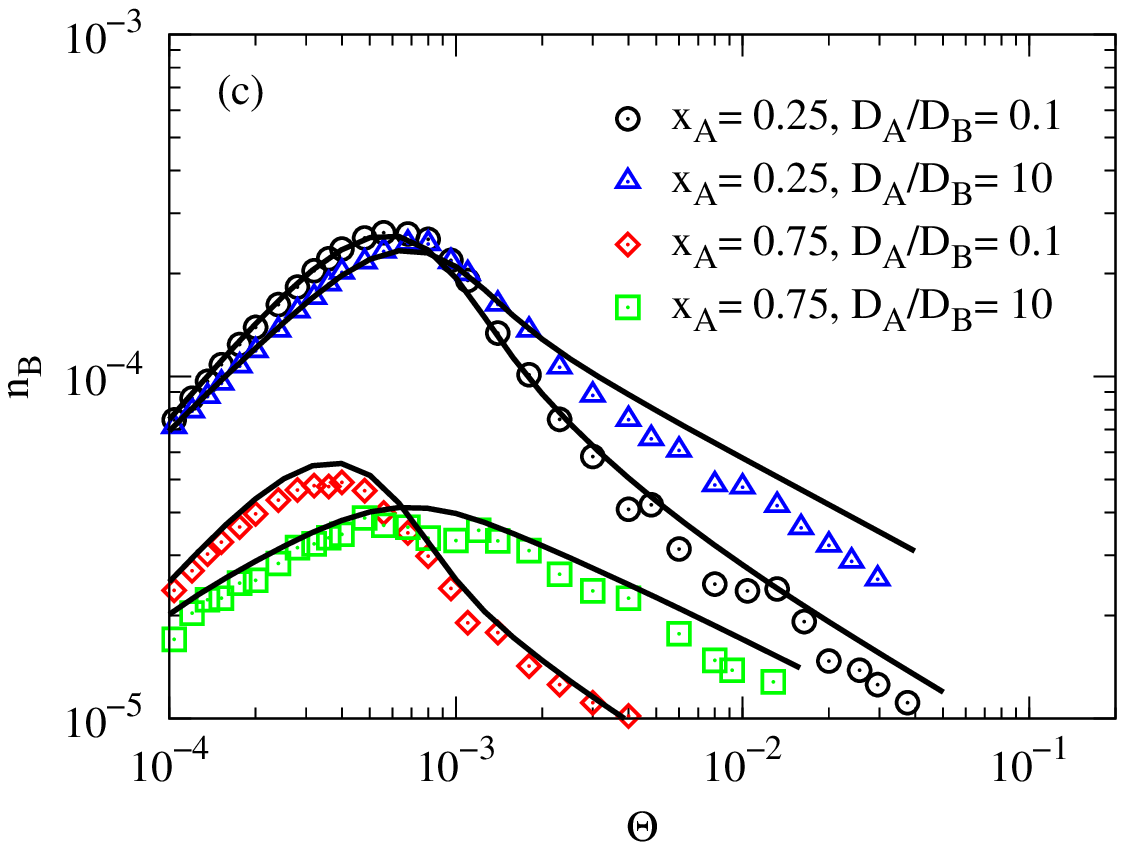}
\caption{(Color online) Number densities of (a) stable islands $N$,
  (b) $A$ adatoms, and (c) $B$ adatoms as a function of the coverage
  $\Theta$ for a case of mixed dimer stabilities, where $AA$ and $AB$
  dimers are stable, while $BB$ dimers are unstable with zero binding
  energies. Results from the self-consistent approach (solid lines)
  with $\mu_{BB}=0.1$ are compared with KMC simulations (symbols).}
\label{fig:fig2}
\end{figure}
\begin{figure}[!htbp]
\includegraphics[width=0.4\textwidth,clip=,]{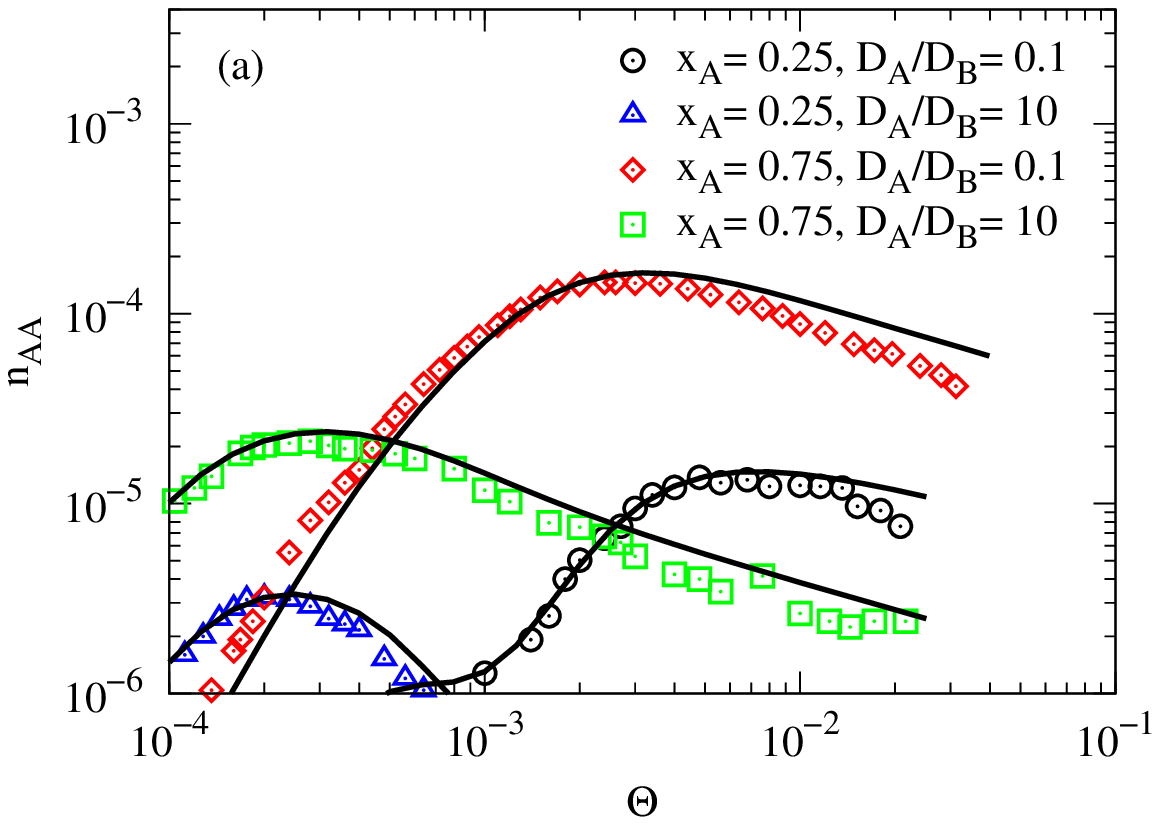}
\includegraphics[width=0.4\textwidth,clip=,]{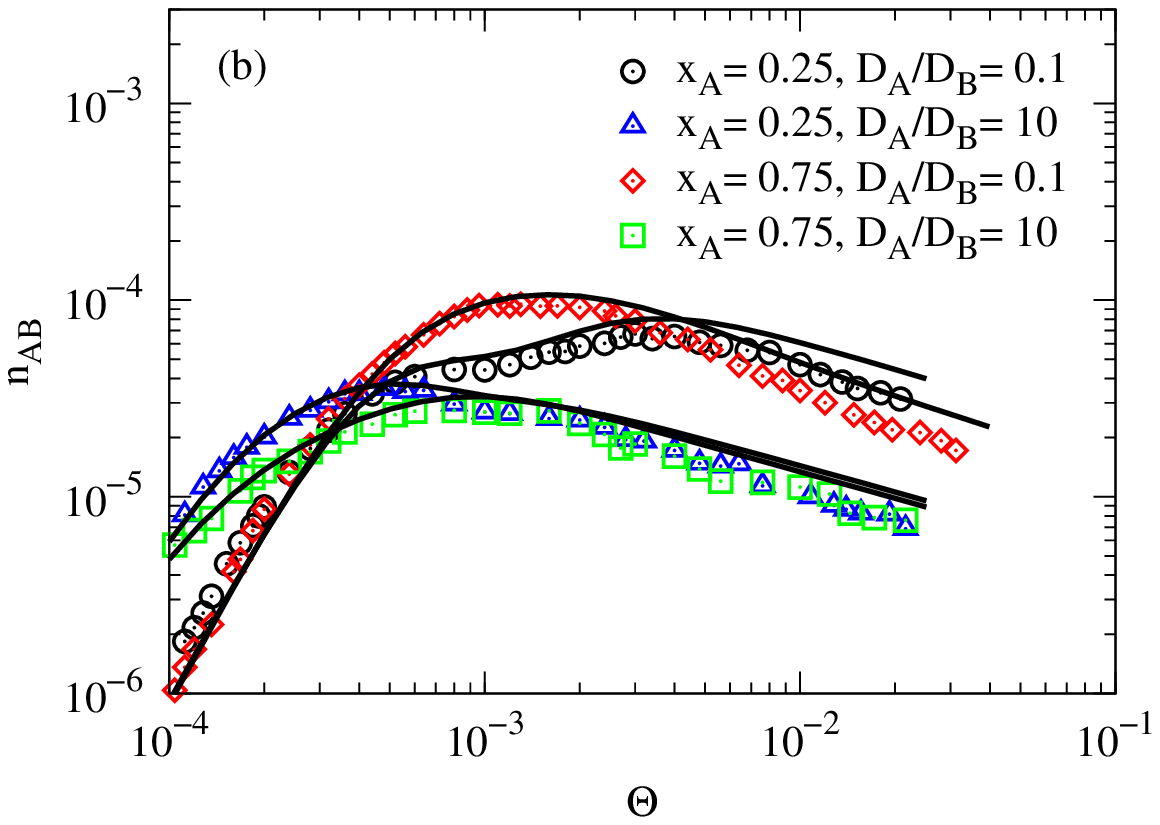}
\includegraphics[width=0.4\textwidth,clip=,]{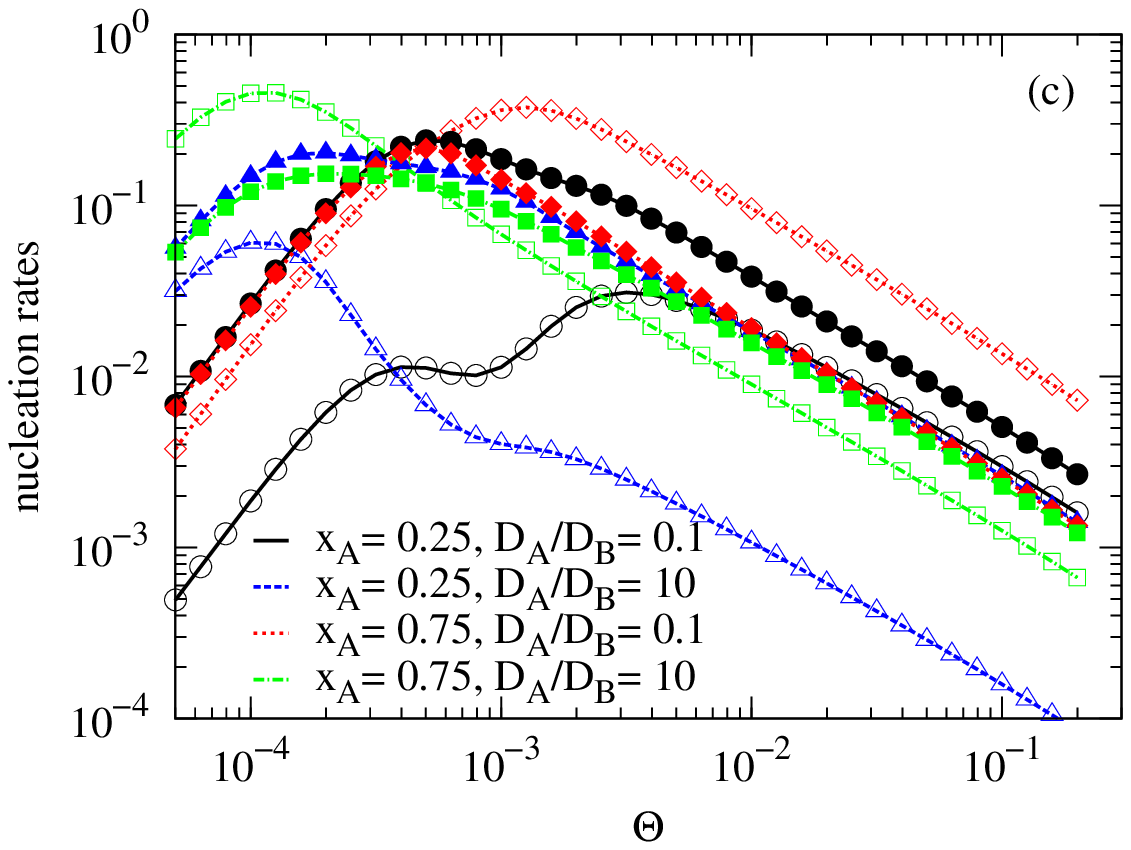}
\caption{(Color online) Densities of (a) stable $AA$ dimers, (b) stable
  $AB$ dimers, and (c) partial nucleation rates $2D_A\sigma_1^{AA}
  (n_A)^2$ (open symbols) and $(D_A+D_B) \sigma_1^{AB} n_An_B$ (filled
  symbols) as a function of the coverage $\Theta$ for mixed dimer
  stabilities as in Fig.~\ref{fig:fig2}.  In (a) and (b) the symbols
  refer to KMC results, and in (c) they are used for the assignment of
  the lines to the parameters. In all figure parts the lines represent
  results of the self-consistent rate theory.}
\label{fig:fig3}
\end{figure}

\section{Numerical results and discussion}
\label{sec:results}
The coupled set of rate equations
(\ref{eq:monomer_densities})-(\ref{eq:s-cluster}) along with the
self-consistent expressions for the capture numbers must be solved
numerically by using an iterative integration scheme. An adequate but
time-consuming numerical integration requires to solve a large number
of equations for an $s$-range in Eqs.~(\ref{eq:s-cluster})
significantly exceeding the mean island size $\bar s=\Theta/N$.  A
much simpler approach of almost the same quality has been proposed for
one-component systems by Venables\cite{Venables:1973} and can be taken
over to binary mixtures considered here. In the case $i=1$, where
$K_2^{\alpha \beta}=0$, this approach amounts to setting
\begin{align}
\label{eq:definition_sigma_av}
\sum_{s\ge 2} \sigma_s^{\alpha}n_s =\bar{\sigma}^{\alpha} N
\end{align}
in Eq.~(\ref{eq:xi_alpha}), to be combined with
Eq.~(\ref{eq:MFRE_with_tau}). Here, $\bar{\sigma}^{\alpha}$ is the
average capture number of stable clusters. Inserting the results from
Sec.~\ref{sec:i1} for $\sigma_s^{\alpha}$ [see the discussion
following Eq.~(\ref{eq:sigma_1_AB})] and assuming that $n_s$ is
sufficiently peaked around the mean island size $\bar{s}$,
one obtains
\begin{align}
\label{eq:averaged_sigma}
\bar{\sigma}^\alpha &= 2 \pi \frac{\bar{R}}{\xi_{\alpha}}
\frac{\mathcal{K}_1 \left( \bar{R}/\xi_{\alpha} \right)}{\mathcal{K}_0
  \left(\bar{R}/\xi_{\alpha} \right) } \, ,
\end{align}
where $\bar{R}=R_{\bar s}=(\bar{s}^{1/2}+1)\mathcal{R}_1$.  The
self-consistency problem then reduces to solving three coupled
equations, Eq.~(\ref{eq:MFRE_with_tau}) for $\alpha=A$ and $B$, and the
equation for nucleation of stable clusters,
\begin{align}
\label{eq:N_reduced_i1}
\frac{dN}{dt} &= \sum_\alpha D_\alpha \sigma_1^{\alpha \alpha}
n_\alpha^2 + (D_A + D_B) \sigma_{1}^{AB} n_A n_B\, .
\end{align}
Capture numbers $\sigma_1^{\alpha \beta}$ and $\bar{\sigma}^{\alpha}$
entering these equations become functions of $n_{\alpha}$ and $N$.

In the more general case of Sec.~\ref{sec:decay}, allowing dimer decay
processes, we must distinguish between stable and unstable dimers. The
example considered below refers to unstable $BB$ dimers but stable
$AA$ and $AB$ dimers, which entails the decomposition
\begin{align}
\label{eq:subst_BB}
\sum_{s\ge 2} \sigma_s^{\alpha} n_s =\sigma_2^\alpha n_{BB}+
\overline{\sigma}^\alpha N \, .
\end{align}
The relevant rate equations now include Eq.~(\ref{eq:Dimer_densities})
for $\alpha=B$ and
\begin{align}
\label{eq:N_reduced_BBunstable}
\frac{dN}{dt} &= D_A \sigma_1^{AA} n_A^2 + (D_A + D_B) \sigma_{1}^{AB}
n_An_B
\nonumber \\
& + (D_A \sigma_2^A n_A + D_B \sigma_2^B n_B) n_{BB}
\end{align}
instead of (\ref{eq:N_reduced_i1}).

To test the self-consistent theory based on that reduced set of
coupled rate equations, we have performed KMC simulations
for codeposition of $A$ and $B$ atoms onto a triangular lattice with
$500\times 500$ sites at various compositions and $D_{A}/D_{B}$
ratios, and for different situations of cluster stabilities with
respect to their size and composition.  Atoms are deposited at random
to vacant substrate sites and diffuse via nearest-neighbour hops,
excluding multiple site occupation.  Attachment of monomers to islands
is accompanied by instantaneous relaxation to highly coordinated edge
sites, yielding compact cluster structures.  For each parameter set,
the number densities were averaged over $50$ realizations.

First, we study the situation of irreversible growth, $i=1$. Results
for $N$ and $n_{\alpha}$ are plotted in Fig.~\ref{fig:fig1} as a
function of the coverage $\Theta=Ft$ for two concentrations $x_A=0.75$
and $x_A=0.25$. In the simulations for both concentrations,
$D_B/F=10^7$ was fixed, and two values $D_A=10D_B$ and $D_A=0.1D_B$
were considered.  The reduced self-consistent theory without fitting
parameters (solid lines) evidently is in good quantitative agreement
with the KMC simulations (open symbols). At low coverages (short
times), $n_{\alpha}=x_{\alpha}\Theta$, whereas in the scaling regime
(see discussion in Refs.~\onlinecite{Evans/etal:2006,
  Einax/etal:2013}), $n_{\alpha} \simeq
x_{\alpha}F/D_{\alpha}N$.\cite{Einax/etal:2007a, Dieterich/etal:2008}
By going from Fig.~\ref{fig:fig1}(a) to (b), the diffusion coefficient
of the majority component $A$ is lowered by a factor $10^2$, which
explains the fact that $n_A$ gets much larger than $n_B$ and the
corresponding curves do not intersect anymore. Inspection of
Eq.~(\ref{eq:N_reduced_i1}) in turn shows that nucleation of stable
islands in Fig.~\ref{fig:fig1}(a) is mostly due to the second term,
i.~e., nucleation of $AB$ dimers prevails, whereas in
Fig.~\ref{fig:fig1}(b) both $AA$ and $AB$ dimers will appear with
similar densities.  In Fig.~\ref{fig:fig1}(b), $N$ close to saturation
becomes significantly larger than in Fig.~\ref{fig:fig1}(a), which is
consistent with the scaling form $N \propto (\Gamma_{\rm eff})^{-1/3}$
with $\Gamma_{\rm eff}=(\sum_{\alpha} x_{\alpha}
F/D_{\alpha})^{-1}$.\cite{Einax/etal:2007a,Dieterich/etal:2008} For
$x_A=0.25$ [Figs.~\ref{fig:fig1}(c) and (d)] the influence of the
mobility ratio $D_A/D_B$ on $N$ is less pronounced. Nucleation in
Fig.~\ref{fig:fig1}(c) proceeds mainly by formation of $BB$ dimers.

Next we include detachment kinetics. Specifically, we assume that the
stability of dimers depends on their composition: $AA$ and $AB$ dimers
are stable ($K_{2}^{AA} = K_{2}^{AB}=0$), while $BB$ dimers are
unstable with zero binding energy.  The number density of stable
islands is given by $N=n_{AA} + n_{AB} + \sum_{s>2 } n_s$ and its time
evolution obeys Eq.~(\ref{eq:N_reduced_BBunstable}).  Again, numerical
results based on our self-consistent rate equations for mixtures are
in good quantitative agreement with the KMC simulations. This is shown
in Figs.~\ref{fig:fig2}(a)-(c) and Figs.~\ref{fig:fig3}(a),(b) for the
same values of $x_A$ and $D_\alpha$ as considered in
Fig.~\ref{fig:fig1}.

A feature worth noting in Fig.~\ref{fig:fig2}(b) is the occurrence of
a local minimum of $n_A$ as a function of $\Theta$ for $x_A=0.25$ and
$D_A/D_B=0.1$.  It can be understood as follows. For these parameters
and throughout the nucleation regime, $AB$ nucleation is the
dominating process for capture of A-atoms, see Figs.~\ref{fig:fig3}(a)
and (c) below. The reason is that $BB$ dissociation entails a large
number of $B$ adatoms, as can be seen in Fig.~\ref{fig:fig2}(c): The
peak in $n_B$ near $\Theta \simeq 10^{-3}$ is about 2.5 times higher
than the corresponding peak in Fig.~\ref{fig:fig1}(d) in the absence
of dissociation. When, with increasing $\Theta$, the $B$ adatom
density $n_B$ approaches its maximum, $AB$ nucleation becomes strong
enough to overcome the gain of $n_A$ by the external flux $F_A$, hence
$n_A$ gets depleted. Beyond $\Theta \simeq 10^{-3}$, on the other hand,
$n_B$ quickly decreases due to reactions with stable islands so that
$n_A$, after going through a minimum, can increase again through
deposition with $F_A$. Upon further increasing $\Theta$, it passes a
second maximum and finally drops through absorption by stable islands.

Shortly speaking, the consumption of $n_A$ after its first maximum in
Fig.~\ref{fig:fig2}(b) is governed by $AB$ nucleation, and after its
second maximum by attachments to stable islands. The rise of the $A$
adatom density after the minimum is due to missing $B$ adatoms for
$AB$ nucleation and the small $D_A$ value. From this discussion it
should become clear, why the minimum is not seen for the curves with
the larger value $D_A/D_B=10$ (shorter mean time to traverse the mean
free path) or the larger $x_A=0.75$ (smaller mean free path for $AA$
nucleation).

To discuss nucleation rates based on
Eq.~(\ref{eq:N_reduced_BBunstable}) and the self-consistent theory,
note first that in all our examples nucleation of trimers via $BB$
dimers is rare, because $n_{BB}$ is small due to decay processes.
Therefore the last term in Eq.~(\ref{eq:N_reduced_BBunstable}) is
negligible. The remaining two terms, giving the partial rates for
nucleation via $AA$ and $AB$ dimers, are represented in
Fig.~\ref{fig:fig3}(c) by open and filled symbols, respectively.  For
example, for $x_A=0.75$ and $D_A/D_B=0.1$, the term $2D_A\sigma_1^{AA}
(n_A)^2$ (open diamonds) becomes larger than the term $(D_A+D_B)
\sigma_1^{AB} n_A n_B$ (filled diamonds). The formation of stable
islands [open diamonds in Fig.~\ref{fig:fig2}(a)] is therefore caused mostly by
the nucleation path via $AA$ dimers. By contrast, for $x_A=0.25$ and
$D_A/D_B=10$ we observe the opposite scenario [see open and filled
triangles in Fig.~\ref{fig:fig3}(c)], which means that $AB$ nucleation
prevails.  In the remaining two cases in Figs.~\ref{fig:fig2} and
\ref{fig:fig3}, both the $AA$ and the $AB$ dimer route contribute
with similar strength to the formation of stable islands.

\section{Conclusions}
\label{sec:conclusion}
We have shown that a self-consistent treatment of capture numbers in
the rate equations for surface growth of binary systems yields a very
good quantitative description of island and adatom
densities. Essential for this theory is the effective absorption
length $\xi_{\rm eff}$ in Eq.~(\ref{eq:effective_length}), which is
symmetric in the two components $A$ and $B$. Its derivation requires
the introduction of pair densities. Note that the weighting factors
$D_{\alpha}/(D_A+D_B)$ appearing in that equation can strongly vary
with temperature as the underlying activation energies for the two
species generally differ. By this, $\xi_{\rm eff}$ acquires an
additional temperature dependence which we expect to become important
in measurements of island and adatom densities.

Different scenarios for dimer stabilities and prevailing nucleation
routes were studied. In all cases, only a reduced set of few coupled
rate equations needs to be solved, which can easily be done on a PC.

Extensions of our theoretical treatment to larger unstable clusters is
straightforward by first generalizing the rate equations as described
in Ref.~\onlinecite{Einax/etal:2007a}. Reduced sets of coupled rate
equations comprise the densities of stable islands, monomers and all
unstable clusters. Extensions to systems with more than two components
and (as before) pairwise reactions follows directly from the above
scheme by introducing pair densities $G_{\alpha \beta}(\mathbf{r})$
among all mobile adatom species $\alpha$ and associated effective
absorption lengths $\xi_{\alpha \beta}$. More generally, in the case
of non-vanishing cluster mobilities,\cite{Jensen/etal:2003} pair
densities need to be introduced for all pairs of mobile species.

\bibliographystyle{apsrev4-1}
\bibliography{einax-etal-lit}

\end{document}